\documentclass{pasj00}

\begin{document}
\SetRunningHead{M. Kubo et al.}{Formation of MMFs and Penumbal Magnetic Fields}
\Received{2007/05/31}
\Accepted{2007/09/02}

\title{Formation of Moving Magnetic Features and Penumbral Magnetic Fields
with Hinode/SOT}

\author{Masahito \textsc{Kubo},\altaffilmark{1,2}
Kiyoshi \textsc{Ichimoto},\altaffilmark{3}
Toshifumi \textsc{Shimizu},\altaffilmark{2}
Saku \textsc{Tsuneta},\altaffilmark{3}
Yoshinori \textsc{Suematsu},\altaffilmark{3}
Yukio \textsc{Katsukawa},\altaffilmark{3}
Shin'ichi \textsc{Nagata},\altaffilmark{4}
Theodore D \textsc{Tarbell},\altaffilmark{5}
Richard A \textsc{Shine},\altaffilmark{5}
Alan M \textsc{Title},\altaffilmark{5}
Zoe A \textsc{Frank},\altaffilmark{5}
Bruce W \textsc{Lites},\altaffilmark{1}
and 
David \textsc{Elmore},\altaffilmark{1}}
 
\altaffiltext{1}{High Altitude Observatory, National Center for
Atmospheric Research\thanks{The National Center for Atmospheric Research
is sponsored by the National Science Foundation.}, P.O. Box 3000,
Boulder, CO 80307, United States} 
\email{kubo@ucar.edu}
\altaffiltext{2}{Institute of Space and Astronautical Science, Japan
Aerospace Exploration Agency, 3-1-1 Yoshinodai, Sagamihara, Kanagawa
229-8510}
\altaffiltext{3}{National Astronomical Observatory of Japan, 2-21-1
Osawa, Mitaka, Tokyo 181-8588}
\altaffiltext{4}{Hida Observatory, Kyoto University, Kamitakara, Gifu 506-1314}
\altaffiltext{5}{Lockheed Martin Advanced Technology Center, O/ADBS, B/252 3251 Hanover Street, Palo Alto, CA 94304 United States}


\maketitle

\begin{abstract}
Vector magnetic fields of moving magnetic features (MMFs) are well
 observed with the Solar Optical Telescope (SOT) aboard the Hinode satellite.
We focus on the evolution of three MMFs with the SOT in this study. 
We found that an MMF having relatively vertical fields with polarity same
 as the sunspot is detached from the penumbra around the granules
 appeared in the outer penumbra. 
This suggests that granular motions in the outer penumbra are
 responsible for the disintegration of the sunspot.
Two MMFs with polarity opposite to the sunspot are located around the
 outer edge of horizontal fields extending from the penumbra.
This is an evidence that the MMFs with polarity opposite to the sunspot are
 prolongation of penumbral horizontal fields.
Radshifts larger than sonic velocity in the photosphere are detected for
 some of the MMFs with polarity opposite to the sunspot.
\end{abstract}

\section{Introduction}
A lot of small magnetic features moving outward are observed in moat
regions surrounding mature sunspots.  
Such magnetic features called moving magnetic features (MMFs,
\cite{Harvey1973}) have been studied for about 40 years from the
discovery by \citet{Sheeley1969}, but their formation process and magnetic
field structure are still open issues due to a difficulty of tracking the
evolution of such tiny magnetic features using vector magnetograms
with high spatial resolution.

The relationship between MMFs and penumbral fields is important to
understand the formation of MMFs and the decaying process of
sunspots.
The penumbral outer boundary has at least two magnetic components
(horizontal fields and relatively vertical fields) called uncombed
structure (\cite{Solanki1993}), and Evershed flows are observed in the
horizontal component of the penumbra(\cite{Degenhardt1991, Title1993,
Lites1993, Rimmele1995, Stanchfield1997, Westendorp2001a,
Westendorp2001b, Mathew2003, Bellot2003, Bellot2004, Ichimoto2007a}).
Some authors have suggested that such Evershed flows drive
outward motions of MMFs (\cite{Ryutova1998, Martinez2002, Thomas2002,
Schlichenmaier2002, Zhang2003}). 
\citet{Sainz2005} observed that several bipolar MMFs crossed the
penumbral outer boundary to enter the moat region along penumbral
horizontal fields. 
\citet{Cabrera2006} discovered that at least some MMFs were the
continuation of penumbral Evershed flows into the moat region.  
\citet{Kubo2007} found MMFs had significant magnetic correspondence
with penumbral uncombed structure from observation of vector magnetic
fields with the Advanced Stokes Polarimeter (\cite{Elmore1992}):
MMFs having horizontal fields with both polarities are located on the
line extrapolated from the horizontal component, while
MMFs having relatively vertical fields with polarity same as the sunspot
are located on the line extrapolated from the vertical component (spine).
Unipolar MMFs with the same polarity as the sunspot are less studied so far,
but they would be more important in terms of the sunspot decay.

In this study, we investigate temporal change of magnetic field vector
for MMFs with the Solar Optical Telescope (SOT,
\cite{Tsuneta2007, Suematsu2007, Ichimoto2007b, Shimizu2007a}) aboard
Hinode (\cite{Kosugi2007}). 
The Hinode/SOT provides us the time series of vector magnetic fields
with the finest resolution (0''.3) ever achieved in the uniform data quality.

\section{Observation and Data Analysis}
\begin{figure}
  \begin{center}
    \FigureFile(80mm,80mm){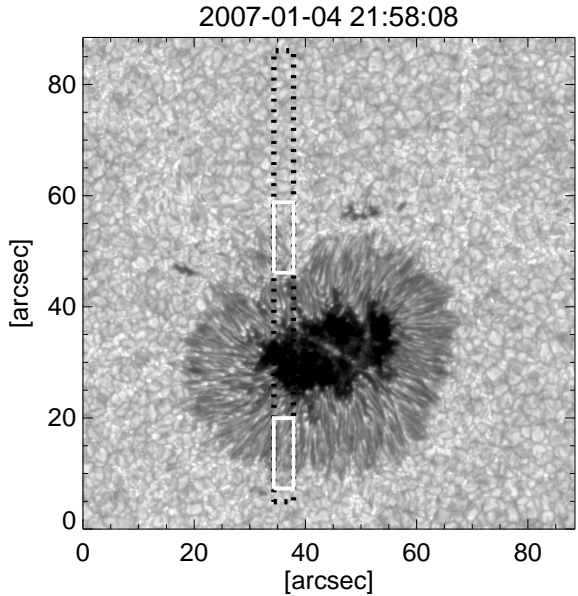}
  \end{center}
  \caption{G-band image with Hinode/SOT for a sunspot in active region
 NOAA 10930 on 2007 January 4. The field of view of the Spectro-Polarimeter
 of the SOT is shown by the dotted box. Two white solid boxes indicate the
 field of view for Figure \ref{fig:mmf_vmmf} (lower) and
 \ref{fig:mmf_hmmf} (upper) respectively.}\label{fig:mmf_Gband}   
\end{figure}

The SOT continuously observed a sunspot in an active region NOAA 10930 during
2007 January 4 to 12 (Figure \ref{fig:mmf_Gband}). 
We used a data set with the Spectro-Polarimeter (SP) of the SOT from
21:57:03 to 22:55:00 in 2007 January 4 when the sunspot was located near
the disk center (S04, E11). 
The SOT/SP repeatedly scanned the same region with a cadence of 128
seconds. 
The slit scanning step was 0$\arcsec$.16, the number of steps in 1 scan is 24
(3$\arcsec$.8), and integration time for one slit position was about 4.8 s. 
The field of view along the slit was 81$\arcsec$ with the pixel
sampling of 0$\arcsec$.16. 
The SOT/SP simultaneously measured the full Stokes profiles (I, Q, U, V)
of the Fe I lines at 630.15 and 630.25 nm with the sampling of 21.6 m{\AA}.


Vector magnetic fields were derived from the calibrated Stokes profiles
on the assumption of Milne-Eddington atmospheres (\cite{Yokoyama2007}). 
We examined field strength($|{\bf{B}}|$), inclination angle($\gamma$)
with respect to the line of sight direction, filling factor($f$), and
Doppler velocity ($v_{los}$) in this paper. 
We show vector magnetic fields by longitudinal field strength
($f|{\bf{B}}|\cos(\gamma)$) and transverse field strength
($f^{1/2}|{\bf{B}}|\sin(\gamma)$). 
The reference of the Doppler velocities was the line center position of Fe I
6302.5{\AA} profiles averaged over the quiet area within the SP map.

\section{Results}
We focus on temporal change of magnetic and velocity fields while MMFs
are formed around the outer edge of the penumbra.
Based on the result of \citet{Kubo2007}, we investigate evolution of
MMFs located on the lines extrapolated from the vertical and horizontal
components of the penumbra separately.

\subsection{MMF detached from the penumbral spine}
\begin{figure*}
  \begin{center}
    \FigureFile(160mm,190mm){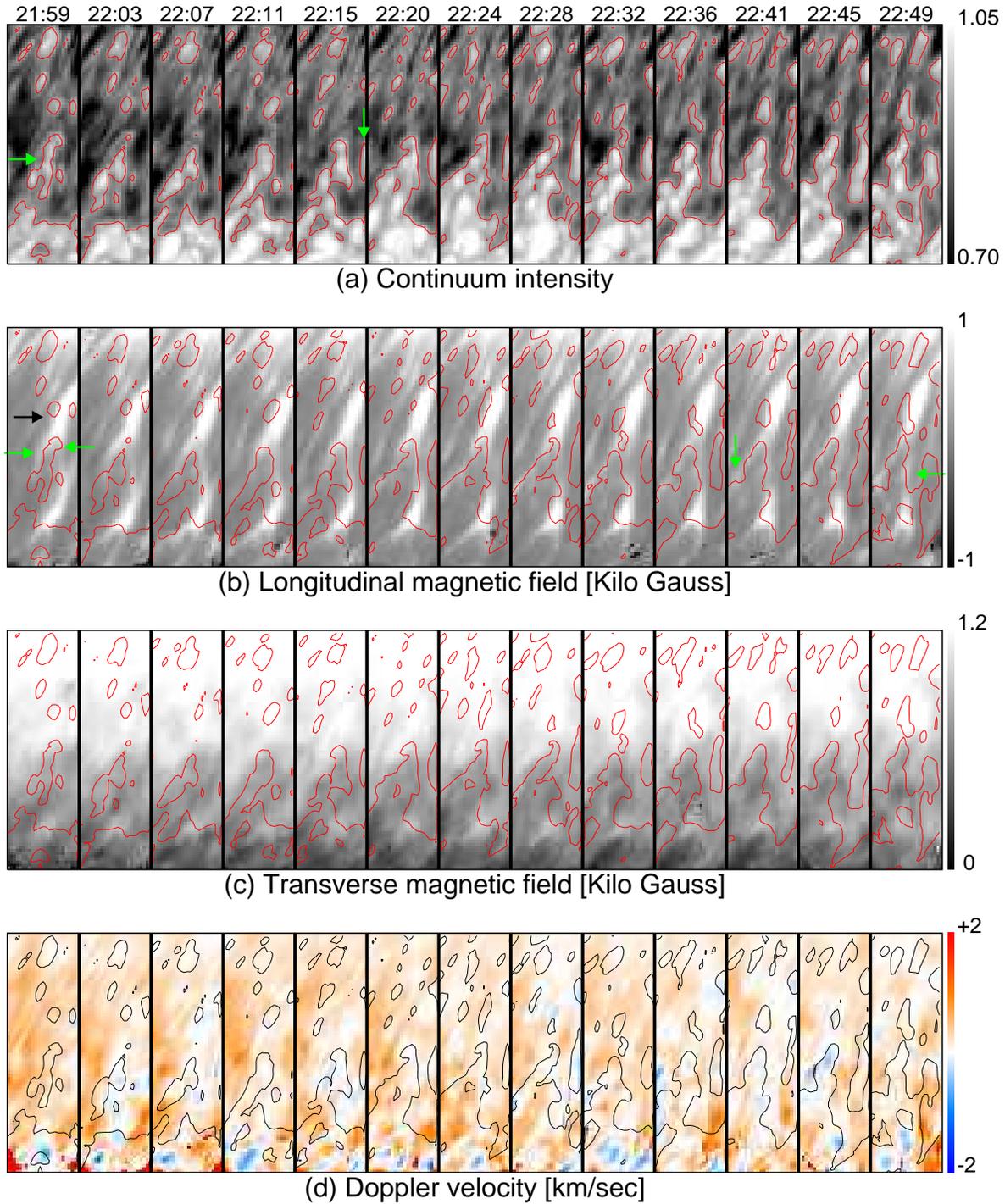}
  \end{center}
  \caption{Time series of (a) continuum intensity, (b) longitudinal
 magnetic field, (c) transverse magnetic field, and (d) Doppler velocity
for MMFs having relatively vertical fields with polarity same as the 
sunspot. The continuum intensity is normalized to the quiet area
 intensity. The contours represent the continuum intensity of 0.87, which
 corresponds to the boundary between the quiet area and the penumbra. White
 (black) is positive (negative) polarity in panel b and positive
 velocity corresponds to redshift.
See the text for arrows in the panels.
}\label{fig:mmf_vmmf}  
\end{figure*}

The penumbral spine is a radial filamentary structure with more vertical
magnetic fields than the surroundings (\cite{Lites1993}), as shown by
the black arrow in Figure \ref{fig:mmf_vmmf} (b).
We find that the penumbral spine branches off at their outer edge (two
green arrows in the first panel of Figure \ref{fig:mmf_vmmf} (b)).
The right branch of the spine is connected to a large magnetic feature
with vertical fields in the first frame and this large magnetic feature
also branches off at its outer edge.  
Then, this large magnetic feature is detached from the penumbral spine
in the later frames. 
We confirmed that the detached magnetic feature moved through the moat
region as the MMF after the SP observation using longitudinal
magnetograms with the Narrowband Filter Imager (NFI) of the SOT.
There is no magnetic feature connected to the left branch of the spine
in the first frame. 
The left branch elongates and a tiny magnetic feature (the green arrow at
22:41 frame in Figure \ref{fig:mmf_vmmf} (b)) subsequently appears
at the outer edge of the elongated branch. 
Such tiny magnetic feature is also formed for the right branch immediately
after the large magnetic feature is detached (the green arrow in the last
frame of Figure \ref{fig:mmf_vmmf} (b)).

We find that the branch of penumbral spine is formed along the
edge of bright features in the penumbra (the horizontal
arrow in Figure \ref{fig:mmf_vmmf} (a)).
The bright features are located around the penumbral spine, but the
penumbral spine itself is not bright.
Such bright features gradually connect to the granular cells outside the
penumbra, and then dark penumbral area between them completely disappears.
Another large bright feature (the vertical arrow in Figure
\ref{fig:mmf_vmmf} (a) at 22:15) appears at the right side of the right
branch while the large magnetic feature with more vertical fields is
separating from the spine. 
The bright features in the penumbra have upward motions and weaker field
strength for both longitudinal and transverse components in comparison
with the surrounding penumbral fields.
These properties are similar to granular cells outside the penumbra. 
The difference in magnetic and velocity fields from the surrounding penumbra
is less clear for the bright features in the inner penumbra. 

We can see a lot of velocity features other than upward motions of the
bright features, but there is no systematic flow related with the
formation of penumbral spine and MMFs detached from the spine. 

\subsection{MMFs located on the lines extrapolated from the horizontal
component of the penumbra}
\begin{figure*}
  \begin{center}
    \FigureFile(160mm,160mm){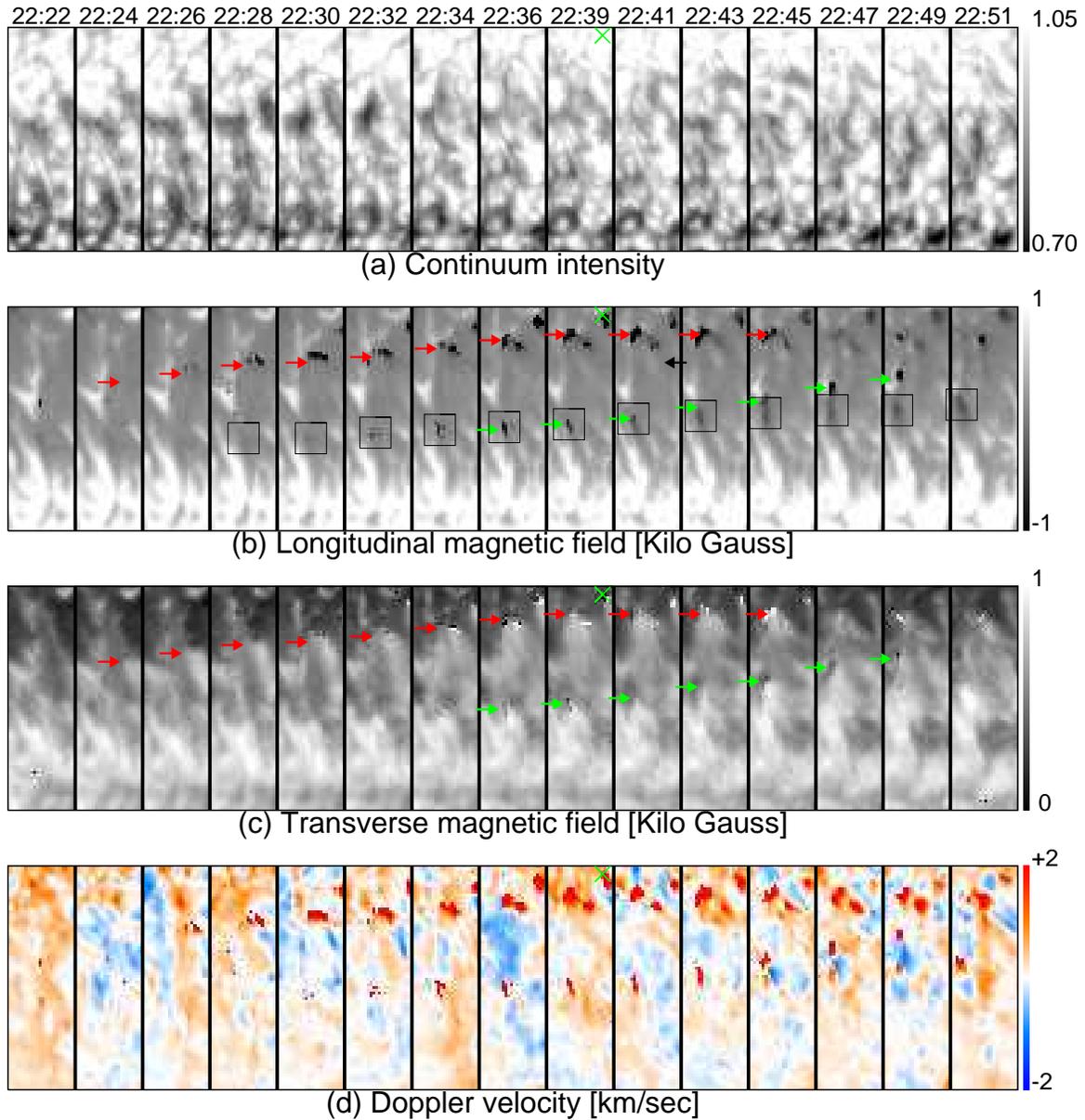}
  \end{center}
  \caption{Same as Figure~\ref{fig:mmf_vmmf} but for MMFs with polarity
 opposite to the sunspot. 
The red and green arrows in panels b and c represent the outer edges of
 horizontal fields extending from the penumbra, and the black arrow 
 represents a position of the MMF with positive polarity.
The cross symbol indicates a position for which the Stokes profiles are
 displayed in Figure \ref{fig:hmmf_profile1}.
The center of boxes in panel b are manually determined to track the
 center of a MMF with negative polarity. 
The first box is located at the same position as the second one because
 the MMF appears from the second box.
}\label{fig:mmf_hmmf}
\end{figure*}

Only MMFs with negative polarity, which is polarity opposite
to the sunspot, are described here in detail because they are clearly
identified from their surroundings.
We discover that two MMFs with polarity opposite to the sunspot appear at
the outer edge of horizontal fields extending from the penumbra, and move
outward with the extending horizontal fields. 
The areas with stronger transverse field are extended to the moat region
from the penumbra, as shown by the red and green arrows in Figure
\ref{fig:mmf_hmmf} (c). 
The outer edge of extending horizontal fields clearly correspond to
positions of the two MMFs with negative polarity.
The fuzzy and elongated MMF with positive polarity is observed
inside the extending horizontal fields (the black arrow in Figure
\ref{fig:mmf_hmmf} (b)). 
These horizontal fields extending from the penumbra have been observed
around the outer boundary of the penumbra in other data sets.

\begin{figure*}
  \begin{center}
    \FigureFile(160mm,120mm){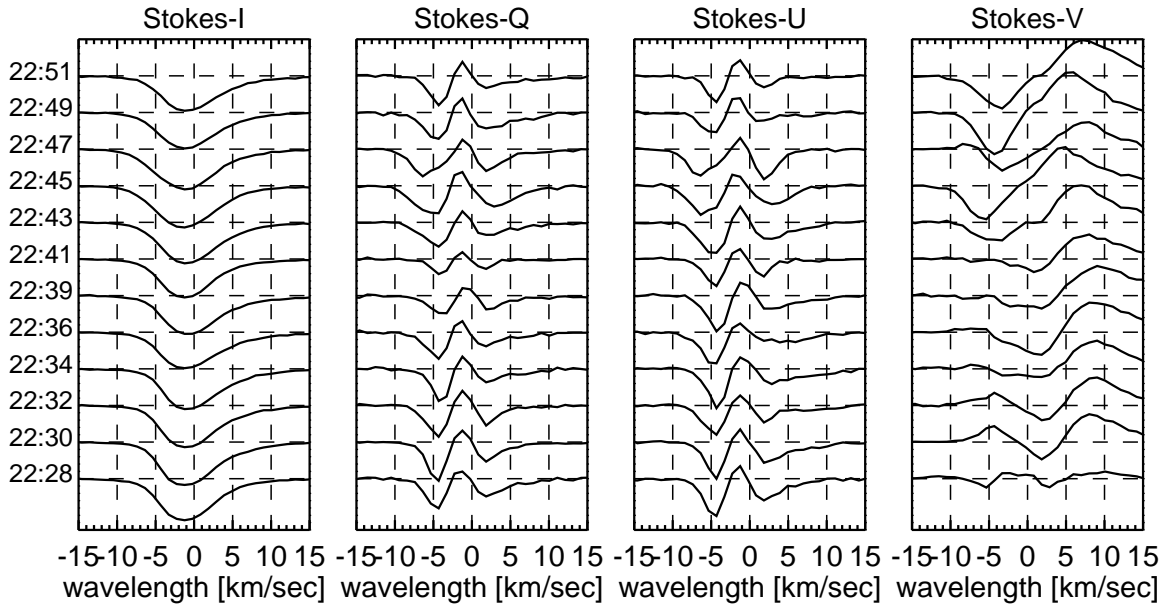}
  \end{center}
  \caption{Time series of Stokes profiles at the center of the boxes in
 Figure \ref{fig:mmf_hmmf}.
Stokes Q, U, and V are shown with the same vertical scale, and the scale
 of Stokes I is 10 times larger than the other profiles.
Zero velocity of horizontal axis corresponds averaged velocity over the
 quiet area. The MMF appears at 22:30.}\label{fig:hmmf_profile2} 
\end{figure*}

Figure \ref{fig:mmf_hmmf} (b) and (d) show that the negative polarity
MMFs have large redshifts.
This result has been already mentioned in previous works
(\cite{Westendorp2001a}, \cite{Cabrera2006}).
One example of evolution in Stokes profiles is shown in Figure
\ref{fig:hmmf_profile2}. 
We find that the shift of linear polarization signals (Q and U) profiles
are much different from that of circular polarization (V):
a major component of the linear polarization has a blueshift of about -1
km/s, while a major component of the circular polarization has a
redshift of about +5 km/s when the negative MMF appears.
Note that the large downward component is seen not only Stokes V profile
but also in Stokes Q and U profiles: the Stokes Q and U have an extended
wing that reaches the range of 5$-$10 km/s. 
The major component of the linear polarization does not significantly
change with time, while the redshifted circular polarization decreases
with time.  
The duration of the large redshift is more than 10 minutes for the two
cases in Figure \ref{fig:mmf_hmmf}, and \citet{Shimizu2007b} shows a
typical duration of such large redshift around the penumbrae is 5$-$20
minutes. 

\begin{figure*}
  \begin{center}
    \FigureFile(160mm,40mm){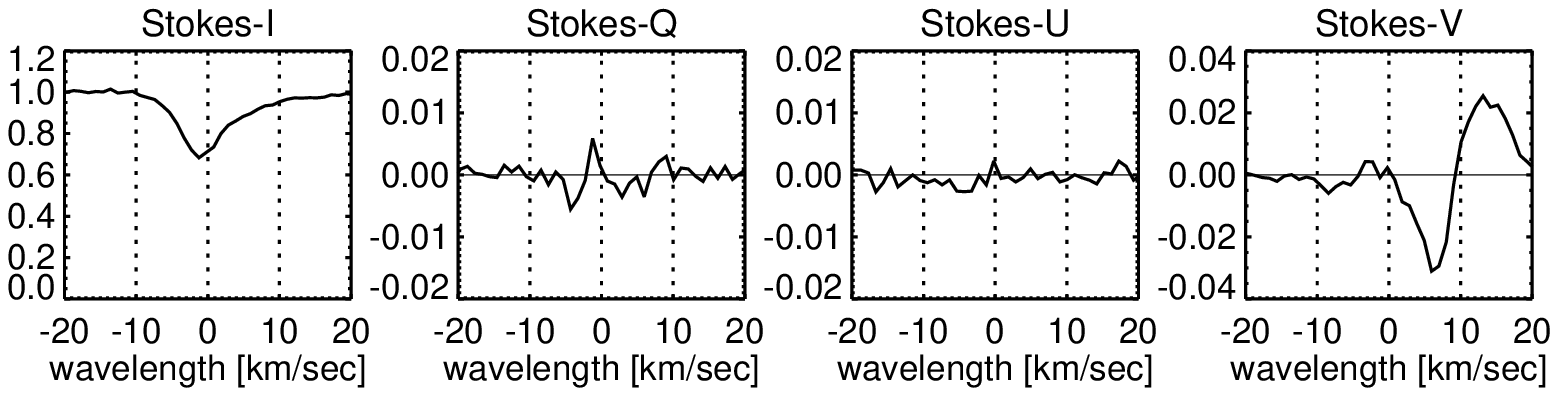}
  \end{center}
  \caption{An example of Stokes profiles for the negative polarity MMF
 (the cross symbol in Figure \ref{fig:mmf_hmmf}).
Vertical axis is normalize by a continuum intensity.
Zero velocity of horizontal axis corresponds averaged velocity over the
 quiet area.}\label{fig:hmmf_profile1}  
\end{figure*}

The redshifts larger than sonic velocity ($\sim$ 7 km/s) in the photosphere
are detected for some of the MMFs with negative polarity in this study,
as shown in Figure \ref{fig:hmmf_profile1}. 
Zero-cross position of Stokes V profile is estimated to be about +9 km/s.
In this case, significant linear polarization signals(Q and U) are not
observed, but linear polarization is often observed in the MMFs with
negative polarity as well as the circular polarization, as shown in
Figure \ref{fig:hmmf_profile2}.

\section{Discussion}
\begin{figure}
  \begin{center}
    \FigureFile(80mm,40mm){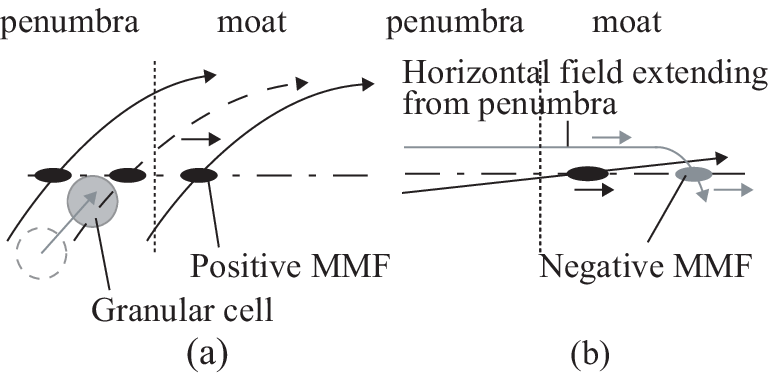}
  \end{center}
  \caption{Schematic illustration for (a) the MMFs detached from the
 penumbral spine and (b) the MMFs located around horizontal fields extending
 from the penumbra.}\label{fig:mmf_model} 
\end{figure}

Our observational result of the MMF detached from the penumbral spine
shows that the branching and elongating of the spine are the basic
process of the formation of the vertical MMF. 
Such vertical MMFs can carry away sufficient magnetic flux causing the
loss of the sunspot magnetic fields (\cite{Kubo2007}).
Therefore, we suggest that granules appeared in the outer penumbra
remove the vertical MMF from the penumbra and such granules are
responsible for the decay of the sunspot (Figure \ref{fig:mmf_model} (a)).   
We believe that the observed MMF is a representative case of the
vertical MMFs with polarity same as the sunspot, but the branches of the
penumbral spines and the granules in the outer penumbra are not observed
for all the spines. 
Many small spines themselves seem to be elongating rather than branching.
Whether the granules appear in the outer penumbra may depend on the size and
amount of field-free gas under the photospheric surface.
It is necessary to confirm the nature of the observed MMF is common to
other vertical MMFs.

The fact that the MMFs with polarity opposite to the sunspot are located
at the outer edge of horizontal fields extending from the penumbra is
an evidence that such MMFs are continuation of the penumbral fields.
This result supports the idea that such MMFs and the extending
horizontal fields have the shape of a sea serpent (\cite{Harvey1973,
Schlichenmaier2002}). 
The MMFs with both polarities located at the outer edge and the inside of the
extending horizontal fields are intersections with the photospheric
surface (formation layer of Fe I lines).
The extending horizontal fields correspond to diffuse and ubiquitous
moving moat fields in previous observations (\cite{Sainz2005},
\cite{Kubo2007}).
The SOT, for the first time, clearly resolves the diffuse structure in the
moat region observed with lower spatial resolution in the past and
reveals the moving moat fields extending from the penumbra are
dynamical phenomena.

The complicated Stokes profiles for the MMFs with opposite polarities to
that of the sunspot are reported in \citet{Cabrera2006}, and the
complication of profiles seems to increase in the SOT data due to the
resolution of magnetic fields with large velocity from ambient fields
without a motion. 
The assumption of Milne-Eddington atmosphere is too simple for
quantitative discussion on the MMFs having the Stokes V profile with
more than two lobes.
However, the MMFs with polarity opposite to the sunspot apparently have
the larger circular polarization signal than the linear polarization in
the red wing.
This means that such MMFs have more vertical than the horizontal fields
extending from the penumbra and large downward motion in the deep layer
of photosphere.
The extending horizontal fields drastically change at the patchy area of
MMFs with polarity opposite to the sunspot (Figure \ref{fig:mmf_model} (b)).

\bigskip
Hinode is a Japanese mission developed and launched by ISAS/JAXA, with
NAOJ as domestic partner and NASA and STFC (UK) as international
partners. It is operated by these agencies in co-operation with ESA and
NSC (Norway).
The authors are grateful to Prof. T. Kosugi and all people who
contributed to this mission.
This work was partly carried out at the NAOJ Hinode Science Center,
which is supported by the Grant-in-Aid for Creative Scientific Research
The Basic Study of Space Weather Prediction from MEXT, Japan (Head
Investigator: K. Shibata), generous donations from Sun Microsystems, and
NAOJ internal funding.

\end{document}